\documentclass[aps,pre,twocolumn,amsmath,amssymb,showpacs]{revtex4-1}
\usepackage{amsmath}
\usepackage{amssymb}
\usepackage{graphicx}
\usepackage{float}
\usepackage{subfigure}
\usepackage{color}
\usepackage{comment}
\usepackage[utf8x]{inputenc}
\usepackage{tabularx}
\usepackage{array}

\begin{document}
\title{Husimi lattice solutions and the coherent-anomaly-method analysis for hard-square lattice gases}
\author{Nathann T. Rodrigues}
\email{nathan.rodrigues@ufv.br}
\author{Tiago J. Oliveira}
\email{tiago@ufv.br}
\affiliation{Departamento de Física, Universidade Federal de Viçosa, 36570-900, Vi\c cosa, MG, Brazil}
\date{\today}

\begin{abstract}
Although lattice gases composed by $k$NN particles, forbidding up to their $k$th nearest neighbors of being occupied, have been widely investigated in literature, the location and the universality class of the fluid-columnar transition in the 2NN model on the square lattice is still a topic of debate. Here, we present grand-canonical solutions of this model on Husimi lattices built with diagonal square lattices, with $2L(L+1)$ sites, for $L \leqslant 7$. The systematic sequence of mean-field solutions confirms the existence of a continuous transition in this system and extrapolations of the critical chemical potential $\mu_{2,c}(L)$ and particle density $\rho_{2,c}(L)$ to $L \rightarrow \infty$ yield estimates of these quantities in close agreement with previous results for the 2NN model on the square lattice. To confirm the reliability of this approach we employ it also for the 1NN model, where very accurate estimates for the critical parameters $\mu_{1,c}$ and $\rho_{1,c}$ --- for the fluid-solid transition in this model on the square lattice --- are found from extrapolations of data for $L \leqslant 6$. The non-classical critical exponents for these transitions are investigated through the coherent anomaly method (CAM), which in the 1NN case yields $\beta$ and $\nu$ differing by at most 6\% from the expected Ising exponents. For the 2NN model, the CAM analysis is somewhat inconclusive, because the exponents sensibly depend on the value of $\mu_{2,c}$ used to calculate them. Notwithstanding, our results suggest that $\beta$ and $\nu$ are considerably larger than the Ashkin-Teller exponents reported in numerical studies of the 2NN system.
\end{abstract}

\maketitle

\section{Introduction}
\label{secIntro}

Entropy-driven phase transitions have received a lot of attention for their role in the packing of dense fluids \cite{BarrySimon}, in granular systems \cite{Granular,Ritort}, in adsorption of molecules onto a surface \cite{JimRSA} and so on. In fact, a large number of studies have considered hard-disks, hard-spheres and other hard-objects, through analytical treatments and numerical simulations, in the continuous space (see, e.g., Refs. \cite{Frenkel,Lafuente2} and references therein). In another front, lattice gases (LGs) composed by hard-particles have been also widely investigated through different methods for a number of particle shapes, such as triangles \cite{Nienhuistri}, dimers \cite{dimers}, rectangles \cite{rectangles}, pentagons \cite{pentagons}, tetrominoes \cite{tetrominoes}, rods \cite{rods}, Y-shaped \cite{RajeshY}, cubes \cite{Rajeshcubes}, etc. on different lattices.

Among these hard-LGs, certainly the most studied ones are those of $k$NN particles, which forbid up to their $k$th nearest neighbor (NN) sites of being occupied by other particles, once they are discrete approximations for hard-spheres and hard-disks and find applications in several areas \cite{Rajesh}. While the $k=0$ case is boring, since the point particles present only a trivial thermodynamic behavior, for $k \geqslant 1$ this simple athermal LGs can present single or multiple transitions from disordered to ordered phases. For instance, on the simple cubic lattice, the 1NN model displays a continuous order-disorder transition \cite{Gaunt,Yamagata,HB,Panagiotopoulos} in the 3D Ising universality class \cite{HB}, while in the 2NN case this transition seems to be discontinuous \cite{Orban2,Panagiotopoulos}, as well as for some larger values of $k$ \cite{Panagiotopoulos}. Results for the body-centered- and face-centered-cubic lattices can be found, e.g., in \cite{Lafuente,Huckaby} and references therein. On the honeycomb lattice, the 1NN model is long known to present a continuous fluid-solid transition belonging to the 2D Ising class \cite{Runnels,Debierre}, whereas only very recently systems for larger $k$'s were analyzed on this lattice \cite{Heitor20}, revealing that the 2NN case presents three stable phases: columnar, solid-like and fluid, with a first-order transition separating the first two ones  \cite{Heitor20}. On the triangular lattice, the 1NN case is Baxter's hard-hexagon model \cite{baxterHH,baxterBook}, the single $k$NN one for which an exact solution is available, showing that it displays a continuous fluid-solid transition in the class of the 3-state Potts model. The 2NN model \cite{Orban,Runnels2,Ted,Zhang} and more recently larger $k$'s \cite{Akimenko,Darjani} have been also investigated on the triangular lattice, where the continuous fluid-solid transition in the 2NN case seems to be in the 4-state Potts class \cite{Ted,Zhang}.

Despite all these works, most of the studies on $k$NN models have been performed on the square lattice, which is also the case of interest here. Since its introduction approximately 70 years ago \cite{DombBurley}, the 1NN hard-square model has been considered in a vast number of works \cite{GauntFisher,Runnels3,Runnels4,Ree,Bellemans,Nisbet,Baxtersql1NN,Binder,Racz,Meirovitch,Pearce,Hu,Hu2,Baram,Jim,GuoBlote,Heitor,Chan,Jensen,Lafuente} and it is well-known to undergo a continuous fluid-solid transition in the 2D Ising class. Particularly in the transfer-matrix study by Guo and Bl\"ote (GB) \cite{GuoBlote}, this was firmly established and the critical chemical potential and particle density were accurately estimated as $\mu_{1,c} = 1.334 015 100 277 74(1)$ and $\rho_{1,c} = 0.367 742 999 041 0(3)$. More recent works have focused on extended hard-core exclusions \cite{Heitor,Rajesh,Rajesh3} --- for instance, $k \leqslant 820 302$ was analyzed in Ref. \cite{Rajesh3} --- and some of them revealed the existence of multiple phase transitions in these systems for large $k$'s \cite{Rajesh,Rajesh3}.

The $2\times 2$ hard-square model (i.e., the 2NN case on the square lattice) has also received much attention in the literature \cite{Binder,Heitor,Bellemans,Bellemans2,Ree2,Nisbet2,Kinzel,Slotte,Amar,Lafuente2,Schimidt,Zhitomirsky,Feng,Ramola,Ramola2,Rajesh2NN,Rajesh2} and it is known to present a disordered fluid and an ordered columnar phase for low and high particle densities, respectively. However, the nature, the location and even the existence of a transition between such phases have been a subject of constant debate. In fact, this is a difficult system, for which different approximation methods usually return quite different outcomes for $\mu_{2,c}$, $\rho_{2,c}$ and the order of the transition (see e.g. the tables in Refs. \cite{Heitor,Rajesh2NN} for summaries of the existing results). While the most recent studies on this model agree that the transition is continuous, different universality classes have been suggested for it. For instance, it was claimed in Ref. \cite{Heitor} that it is the class of the 2D Ising model, similarly to the 1NN case, while exponents close, but deviating from the Ising ones were subsequently reported in \cite{Zhitomirsky,Feng,Ramola2}. Particularly in Ref. \cite{Ramola2}, convincing evidence that this system presents Ashkin-Teller criticality was reported, as previously hinted in \cite{Feng}.

In view of this discussion --- and considering that it is being mainly guided by Monte Carlo (MC) simulations, once other finite-size analysis successfully applied to the 1NN model \cite{GuoBlote} has proved to be inconclusive in the 2NN case \cite{Feng} --- it is important to further investigate the $2\times2$ hard-square model considering other approaches. Here, we address this through semi-analytical grand-canonical solutions on Husimi lattices whose building blocks are diagonal square lattices with $2 L (L+1)$ sites \cite{Monroe}. See Figs. \ref{fig1}, \ref{fig2} and \ref{fig3}. Continuous fluid-columnar transitions are found at all levels of approximation analyzed (up to $L=7$), yielding a series of even better mean-field results for the critical parameters $\mu_{2,c}(L)$ and $\rho_{2,c}(L)$. Extrapolations of these numerically exact critical points to $L\rightarrow \infty$ return values close to the best known estimates for them (from extensive MC simulations on the square lattice). Similarly, by employing the same procedure for the 1NN model, results in quite good agreement with those found by GB \cite{GuoBlote} are obtained for the continuous fluid-solid transition on the square lattice. To investigate the true critical exponents of these systems (for the square lattice), we use the coherent anomaly method (CAM) \cite{SuzukiCAM}. We remark that this method has been applied in the study of the criticality in a diversity of classical and quantum systems, as well as nonequilibrium ones \cite{SuzukiBook,SuzukiChap}. However, to the best of our knowledge, for LGs there exits a single study applying CAM to soft (Lennard-Jones type) systems \cite{Patrykiejew}, where the method has failed in providing the expected critical exponents. So, our work may serve also as a check of the effectiveness of the CAM analysis for hard-LGs and entropy-driven phase transitions. While exponents close to the expected Ising ones are found for the 1NN model, our results do not allow us to draw a firm conclusion on the universality class in the 2NN case.

The outline of this paper is as follows. In Sec. \ref{secModel} we define the $k$NN models and devise the method for solving them on HLs of different levels. Results for the critical parameters for the 1NN and 2NN model are presented in Secs. \ref{secRes1NN} and \ref{secRes2NN}, respectively. In Sec. \ref{secCAM} the CAM analysis is applied to both models. Section \ref{secConc} summarizes our final discussions and conclusions.

\section{Models and methods}
\label{secModel}

\subsection{Models}

\begin{figure}[t]
 \includegraphics[width=8.3cm]{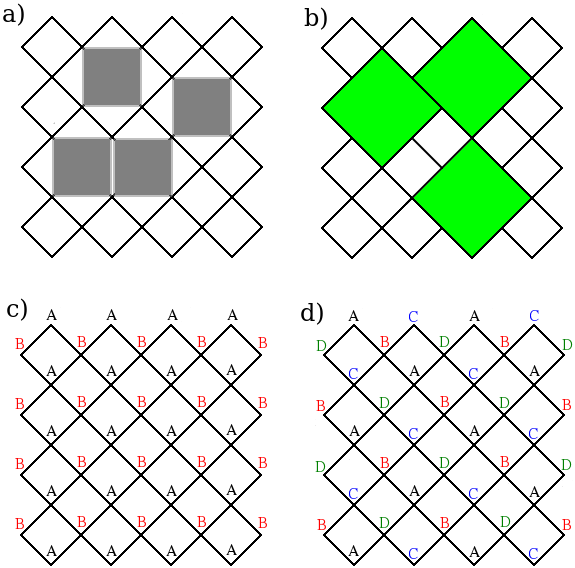}
 \caption{Illustration of a) 1NN and b) 2NN particles on a diagonal square lattice with $L=4$. The definitions of sublattices for studying the 1NN and 2NN models are respectively presented in (c) and (d). %The sites of the colored plaquettes in (c) and (d) are the ones where the particle densities will be calculated.
 }
 \label{fig1}
\end{figure}

\begin{figure*}[t]
 \includegraphics[width=16.5cm]{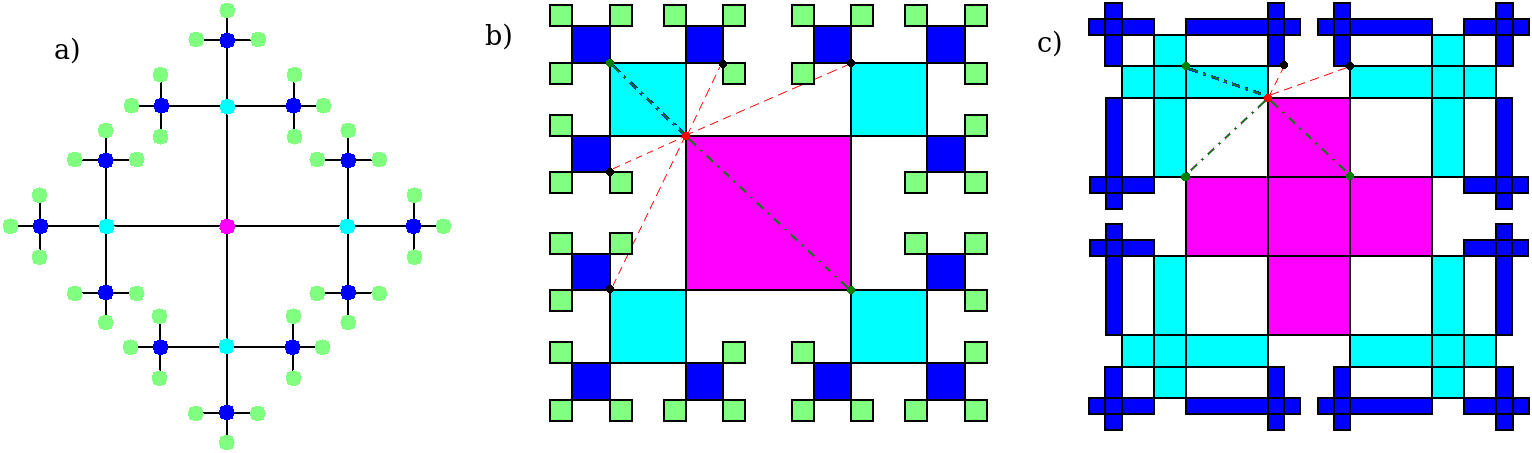}
 \caption{a) Bethe lattice with coordination $q=4$. Square Husimi lattices of b) first- ($L=1$) and c) second-level ($L=2$). Lattices with three [two] generations are shown in (a) and (b) [(c)], once we are considering the central (magenta) building blocks as the starting point. Different colors are associated with each generation. The dashed red lines in (b) and (c) indicate the NNNs \textit{external to the plaquettes} of the red sites they are emanating, while the dash-dotted green lines connect the red sites with their \textit{internal} NNNs.}
 \label{fig2}
\end{figure*}

A given $k$NN model is composed by hard-core particles, placed on (and centered at) the vertices of a given lattice, which exclude up to their first $k$ next nearest neighbors of being occupied by other particles. In our grand-canonical treatment of these systems, an activity $z_k=e^{\mu_k}$, where $\mu_k=\tilde{\mu}_k/k_B T$ is the reduced chemical potential, will be associated with each $k$NN particle. For the sake of simplicity, hereafter we will refer to $\mu_k$ simply as ``the chemical potential''. On the square lattice, the 1NN particles correspond to hard squares of lateral size $\lambda=\sqrt{2}a$ tilted by $45\textdegree$ in relation to the lattice, where $a$ is the lattice spacing [see Fig. \ref{fig1}(a)]. In the full occupancy limit, when $\mu_1 \rightarrow \infty$ and the density is $\rho_{1,max}=1/2$, only one of two sublattices [$A$ or $B$, as defined in Fig. \ref{fig1}(c)] is occupied, so that the system presents long-range order in this solid phase. By decreasing $\mu_1$ a melting transition is observed for a disordered fluid phase, where both sublattices are equally populated (i.e., $\rho_{1A}=\rho_{1B}$). Thereby, an appropriate definition of the order parameter for this transition is \cite{Heitor}
\begin{equation}
 Q_1 = \frac{1}{\rho_{1,max}} |\rho_{1A}-\rho_{1B}|,
 \label{eqQ1}
\end{equation}
since $Q_1 = 0$ ($Q_1 > 0$) in the fluid (solid) phase, being $Q_1=1$ in the ground state. The way to calculate the densities will be devised in the Appendix.

The 2NN particles correspond to hard squares of lateral size $\lambda=2a$ occupying four elementary plaquettes of the square lattice, as is shown in Fig. \ref{fig1}(b). In the limit of $\mu_2 \rightarrow \infty$, where $\rho_2=\rho_{2,max}=1/4$, this system presents a long-range columnar order, rather than a solid phase, due to a sliding instability, and four sublattices [$A,\ldots,D$, see Fig. \ref{fig1}(d)] are needed to characterize this fourfold degenerate ground state. By decreasing $\mu_2$, a transition is expected from the columnar phase to a disordered fluid phase, where $\rho_{2A}=\rho_{2B}=\rho_{2C}= \rho_{2D}$. The fourfold symmetry breaking in such transition can be captured by the order parameter \cite{Heitor}
\begin{equation}
 Q_2 = \frac{1}{\rho_{2,max}} (|\rho_{2A}-\rho_{2C}|+|\rho_{2B}-\rho_{2D}|),
 \label{eqQ2}
\end{equation}
once $Q_2 = 0$ in the fluid phase and $Q_2 > 0$ in the columnar one.

\subsection{Husimi lattice solutions}

A \textit{Bethe lattice} (BL) is the core of an infinite Cayley tree: a hierarchical structure, without loops, which can be built by successively adding $q-1$ edges to each boundary site of the previous generation ($M-1$), starting with a ``central'' site and adding $q$ edges to it to form the first generation of the tree. In this way, all sites in the interior of the tree have coordination $q$, while those at the boundary have a single neighbor [see Fig. \ref{fig2}(a)]. Since loops are absent in the BL, solutions of models on it can be seen as the ``zeroth-level'' (``$L=0$'') treelike mean-field approximation for a given model on a regular lattice. This can be improved by replacing the sites and edges of the BL by clusters, yielding the so-called \textit{Husimi lattices} (HLs) \cite{Husimi}. In the ordinary (first-level) HL approximation for the square lattice, a square cactus is built by connecting neighboring elementary squares by a single vertex, as shown in Fig. \ref{fig2}(b). Similar treelike lattices can be built with triangles, cubes and so on. Recent examples of systems investigated on these Husimi cacti include frustrated magnets \cite{HusimiFM}, polymers \cite{HusimiPol} and lattice gases \cite{HusimiGR}. In particular, quite recently binary \cite{NathannBin} and ternary \cite{NathannTern} mixtures of $k$NN particles were analyzed by us on a HL built with cubes. 

\begin{figure}[t]
 \includegraphics[width=8.cm]{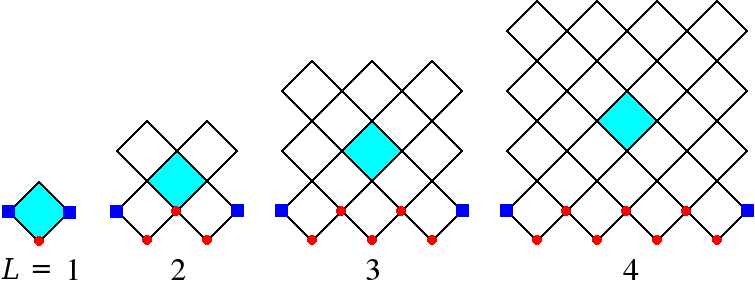}
 \caption{Building blocks of square Husimi lattices of levels $L \leqslant 4$. The generalization for higher $L$'s is immediate. The sites composing the root zigzag lines for the solution of the 1NN model are indicated by the red dots, while for solving the 2NN model the sites indicated by the blue squares are also included in the root lines. The colored plaquettes are the central ones, where the particle densities are investigated.}
 \label{fig3}
\end{figure}

In order to solve a given model on these treelike structures, the symmetries of all of its phases have to be reproduced on such lattices. However, this is not always possible when dealing with the lowest levels. For example, the definition of the 2NN model on the BL is somewhat arbitrary, due to the definition of second neighbors in this lattice. In fact, by doing this considering the chemical distance, a discontinuous order-disorder transitions is obtained for this model \cite{Robledo}. Actually, even in the ordinary square HL of Fig. \ref{fig2}(b), it seems not possible to appropriately account for the correlations of the columnar phase. In such situations, we are compelled to consider higher level HLs, where the elementary plaquette is replaced by a cluster of plaquettes. Here, we will adopt the scheme introduced by Monroe \cite{Monroe}, using diagonal square lattices as building blocks, which share $L$ sites between two consecutive generations of the tree in each of its four sides. Figure \ref{fig3} shows these building blocks for levels $L\leqslant 4$. The HL for $L=2$ is depicted in Fig. \ref{fig2}(c), where some plaquettes had to be deformed to allow the drawing of a tree with more than one generation in the plane.

It is important to remark here that even in HLs the definition of second and higher order neighbors is problematic. Let us concentrate first on the $L=1$ case of Fig. \ref{fig2}(b), where each site clearly has only two next-nearest neighbors (NNNs) internal to the plaquettes, while this number should be four in the square lattice. If one considers also the NNNs external to the plaquettes, each site has four of such neighbors, as indicated by the dashed lines in Fig. \ref{fig2}(b), so that the total number of NNNs now becomes six. As demonstrated in Ref. \cite{tiagoJPA16}, in thermal systems with NNN interactions associated with a Boltzmann weight $\omega$, one way to conciliate this with the square lattice is by associating weights $\omega$ to each of the two internal NNNs and $\sqrt{\omega}$ to each of the four external ones. Since this can not be applied to the athermal 2NN model, we have to either underestimate or overestimate the number of NNN sites. This problem lessens for $L \geqslant 2$, where most of sites have four NNNs internal to the plaquettes. The exceptions are the eight sites at the corners of the building blocks shown in Fig. \ref{fig3}, which have only three internal NNNs, as seen in Fig. \ref{fig2}(c). So, we can solve the 2NN model underestimating (U) the neighborhood of these corner sites, by considering only their three internal NNNs. Another possibility is to consider also the two external NNNs of the corner sites [see Fig. \ref{fig2}(c)], overestimating (O) them. Results for both approaches (O and U) will be presented in the following sections.

The solutions of the $k$NN models on the square HLs are discussed in detail in the Appendix. We anticipate here that such solutions are obtained in terms of recursion relations (RRs) for ratios of partial partition functions (ppf's), which are defined according to the particles' states and sublattice configuration in the root zigzag line of rooted building blocks [see Fig. \ref{fig3}]. These RRs are given by ratios of multivariate polynomials whose number of terms becomes prohibitively large to deal with, even computationally, already for small $L$'s [see the values in the Tab. \ref{tab1}]. This has limited our analysis to $L \leqslant 6$ ($L \leqslant 7$) in the $k=1$ ($k=2$) case. The real, positive and stable fixed points of these RRs define the thermodynamic phases of the models on the HL. Beyond the (reduced) bulk free energies \textit{per site} [$\phi_k = \tilde{\phi}_k/k_B T$, which in our grand-canonical formalism are related to the (reduced) pressure as $P_k=-\phi_k/a^2$], we will study also the particle densities in each sublattice $S$ [$\rho_{kS}$], the total particle densities [$\rho_k = \sum_S \rho_{kS}$] and the order-parameters [$Q_k$, defined in eqs. \ref{eqQ1} and \ref{eqQ2}]. All densities (and then also $Q_k$) will be calculated at the four sites of the central plaquettes of the central building blocks of the HL [see Fig. \ref{fig3}]. These are the sites suffering less with effects from the HLs' boundary, so that results more consistent with the square lattice are expected there \cite{Monroe}.

\section{Critical parameters for the 1NN model}
\label{secRes1NN}

Although our main interest here is in the 2NN model, it is natural to start our analysis with the 1NN case, for which the critical parameters are known with high precision. In all levels considered, $1 \leqslant L \leqslant 6$, two types of fixed points are found for the RRs for the ratios of ppf's, $R_{\sigma,S}$, as defined in the Appendix. One has a fixed point associated with the disordered fluid ($F$) phase, characterized by a homogeneous solution of the RRs, with $R_{\sigma,A}=R_{\sigma,B}$ for $\sigma=1,2,...,N-1$, where $N$ is the total number of ppf's for a given $L$ and sublattice ($A$ or $B$). There are two other equivalent fixed points associated with the ordered solid ($S$) phase, where one sublattice is more populated. For example, $R_{\sigma,A}>R_{\sigma,B}$ when sublattice $A$ is the one more occupied and vice-versa.

The stability analysis, for all levels, reveals that the $F$ ($S$) phase is stable for small (large) $z_1$ and that the spinodals of both phases take place at the same value of $z_1$, which turns out to be a critical point $z_{1,c}$. Therefore, a continuous $F$-$S$ transition is found in the HLs, in agreement with the behavior of the 1NN model on the square lattice. This is confirmed also by the behavior [not shown] of the particle densities (since one finds $\rho_1^{F} = \rho_1^{S}=\rho_{1,c}$, at $z_1=z_{1,c}$), free energies (since one observes that $\phi_1^{F} = \phi_1^{S} = \phi_{1,c}$, at $z_1=z_{1,c}$) and order parameter (since $Q_1 \rightarrow 0$ as $z_1 \rightarrow z_{1,c}$ from above). The values found for the critical parameters $\mu_{1,c}=\ln (z_{1,c})$, $\rho_{1,c}$ and $\phi_{1,c}$, for different $L$'s, are summarized in Tab. \ref{tab2}. For comparison, results for the BL with coordination $q=4$ \cite{tiago11} are also displayed in this table, which can be seen as the ``$L=0$'' case. Since these last values do not follow the systematic convergence observed in the data for $L \geqslant 1$, they will not be used in the extrapolations. Actually, even the results for $L=1$ will be disregarded in the extrapolations below, because they are always quite different from the rest, probably because this is still a very crude approximation for the square lattice.

\begin{table}[b] \centering
\caption{Critical chemical potentials $\mu_{1,c}$, particle densities $\rho_{1,c}$ and free energies $\phi_{1,c}$ for the 1NN model on square HLs of different levels $L$. }
\begin{tabularx}{\columnwidth}{p{0.6cm} >{\centering\arraybackslash}X >{\centering\arraybackslash}X >{\centering\arraybackslash}X >{\centering\arraybackslash}X}%{l c c c c}
 \hline
 \hline
  $L$     & $\mu_{1,c}$   &    $\rho_{1,c}$     &   $\phi_{1,c}$   \\
  \hline
  $0$\footnote{Results for the Bethe lattice with coordination $q=4$ \cite{tiago11}.}     & $0.523248$      &  $0.250000$    &     $-0.261714$         \\
  $1$      & $0.682526$           &    $0.269594$      &  $-0.285784$	\\ 
  $2$      & $0.929908$           &    $0.308815$      &  $-0.433966$	\\
  $3$      & $1.035744$           &    $0.319238$      &  $-0.514850$	\\
  $4$      & $1.096213$           &    $0.327915$      &  $-0.565728$	\\
  $5$      & $1.135595$           &    $0.333132$      &  $-0.600665$	\\
  $6$      & $1.163389$           &    $0.336882$      &  $-0.626140$	\\ 
%   $\infty$\cite{} & $\approx3.79625517$   \\
 \hline
 \hline
\end{tabularx}
\label{tab2}
\end{table}

With the data for the critical chemical potential at hand, we can use different methods for estimating $\mu_{1,c}(L \rightarrow \infty$), which shall provide an estimate of $\mu_{1,c}$ for the model on the infinite square lattice (i.e, in its thermodynamic limit). We start assuming the usual finite-size scaling form
\begin{equation}
 X(L) = X(\infty) + a_1 L^{-\Delta_1} + a_2 L^{-\Delta_2} + ...,
 \label{zc}
\end{equation}
with $X=\mu_{1,c}$, where one expects $0<\Delta_1<\Delta_2<\cdots$. As a first approximation, we can consider that $a_i=0$ for $i \geqslant 2$, letting us with three unknowns [$\mu_{1,c}(\infty)$, $a_1$ and $\Delta_1$], which can be obtained from three-point (3-pt) extrapolations for sets of levels $(L-1,L,L+1)$. The values of $\mu_{1,c}(\infty)$ and $\Delta_1$ estimated in this way are depicted in Tab. \ref{tab1NNExt}. The appreciable variation of these quantities with $L$ indicates that further corrections can not be neglected. For instance, if one performs an additional 3-pt extrapolation of the extrapolated values $\mu_{1,c}(\infty)$ in Tab. \ref{tab1NNExt}, we obtain $\mu_{1,c} \approx 1.3374$, which differs approximately $0.2\%$ from the very accurate value estimated by Guo and Bl\"ote (GB) \cite{GuoBlote}: $\mu_{1,c} \approx 1.3340151002$.

We can improve this by considering also the third term in the rhs of Eq. \ref{zc}, assuming $a_i=0$ only for $i\geqslant 3$. From the exponents in Tab. \ref{tab1NNExt}, it is hard to infer the asymptotic value of  $\Delta_1$. Thereby, we have to perform a 5-pt extrapolation, with $\mu_{1,c}(\infty)$, $a_1$, $a_2$, $\Delta_1$ and $\Delta_2$ as unknowns. This gives $\mu_{1,c}(\infty) \approx 1.33486$ for the largest $L$'s, which is close to the GB value, but with $\Delta_2 \approx \Delta_1 \approx 1$, suggesting the presence of an additive logarithmic term $L^{-\Delta_1}\ln L$ in Eq. \ref{zc}. If one assumes that $\mu_{1,c}(L) = \mu_{1,c}(\infty) + b_1 L^{-\Delta_1} \ln(v_1 L)$, a 4-pt extrapolation for the largest levels yields $\mu_{1,c}(\infty) \approx 1.3337$ and, once again, $\Delta_1 \approx 1$. Furthermore, 5-pt extrapolations considering the existence of the logarithmic term, such as $\mu_{1,c}(L) = \mu_{1,c}(\infty) + c_1 L^{-\Delta_1} + c_2 L^{-\Delta_2} \ln L$ or $\mu_{1,c}(L) = \mu_{1,c}(\infty) + c_1 L^{-1} + c_2 L^{-\Delta_2} (\ln L)^{-\xi_2}$, return $1.3335 \lessapprox \mu_{1,c}(\infty) \lessapprox 1.3345$. Hence, it is reasonable to conclude that the true critical point is located at $\mu_{1,c}^* = 1.3340(5)$, which agrees with and differs from the GB value by $0.001$\%.

To obtain the asymptotic value of the critical density, we can start assuming the finite-size behavior of Eq. \ref{zc} and, then, employing a 3-pt extrapolation to estimate $\rho_{1,c}(\infty)$, $a_1$ and $\Delta_1$. The extrapolated values, $\rho_{1,c}(\infty)$, are summarized in Tab. \ref{tab1NNExt}. Curiously, for the set of levels $(2,3,4)$ no physical solution is found from this extrapolation, indicating that the densities for the smaller $L$'s are not in the convergence regime. In fact, by using them to perform 4- or 5-pt extrapolations, considering pure power-law or logarithmic corrections, a diversity of values are obtained, with $\rho_{1,c}(\infty) \in [0.35,0.39]$. On the other hand, the result from a simple 3-pt extrapolation for the largest levels available [$(4,5,6)$ in Tab. \ref{tab1NNExt}] is quite close to the one by GB ($\rho_{1,c} \approx 0.367742999$ \cite{GuoBlote}), differing by 0.02\% from it.

Similarly to the other quantities, initially, we consider the finite-size scaling of Eq. \ref{zc} also for the free energy $\phi_{1,c}$. The outcomes from 3-pt extrapolations are shown in Tab. \ref{tab1NNExt}. The large variation in these extrapolated values demonstrates that we cannot neglect further corrections in $\phi_{1,c}$. Similarly to $\mu_{1,c}$, a 5-pt extrapolation following Eq. \ref{zc} returns exponents $\Delta_2 \approx \Delta_1 \approx 1$, suggesting the existence of logarithmic corrections. A 4-pt extrapolation assuming that $\phi_{1,c}(L) = \phi_{1,c}(\infty) + b_1 L^{-\Delta_1} \ln(v_1 L)$ for the largest $L$'s yields $\phi_{1,c}(\infty) \approx -0.79447$ with $\Delta_1 \approx 1$, while a 5-pt extrapolation with $\phi_{1,c}(L) = \phi_{1,c}(\infty) + c_1 L^{-1} + c_2 L^{-\Delta_2} (\ln L)^{-\xi_2}$ gives $\phi_{1,c}(\infty) \approx -0.79169$. Once again, these values are in quite good agreement with the one found by GB ($|\phi_{1,c}|=0.791 602 643 166 112(1)$ \cite{GuoBlote}), with a difference of $0.01$\% in the latter case.

\begin{table}[t] \centering
\caption{Results from 3-pt extrapolations of the critical parameters $\mu_{1,c}$, $\rho_{1,c}$ and $\phi_{1,c}$ in Tab. \ref{tab2}, considering Eq. \ref{zc} for sets of levels $(L-1,L,L+1)$. The obtained exponents $\Delta_1$ from the extrapolations of $\mu_{1,c}$ are also shown.}
\begin{tabular}{l c c c c}
 \hline
 \hline
  Set of $L$'s      & \phantom{...}$\mu_{1,c}(\infty)$   &    $\Delta_1$      &   $\rho_{1,c}(\infty)$ &   $\phi_{1,c}(\infty)$  \\
  \hline
%  $(1,2,3)$      & $1.44933$           &    $0.56195$    &  $0.33300$  & $-2.06261$  \\ 
  $(2,3,4)$      & $1.40530$           &    $0.62111$    &     ---     & $-1.05149$  \\
  $(3,4,5)$      & $1.37521$           &    $0.68192$    &  $0.35418$  & $-0.91254$  \\
  $(4,5,6)$      & $1.36141$           &    $0.72041$    &  $0.36766$  & $-0.86300$  \\

%   $\infty$\cite{} & $\approx3.79625517$   \\
 \hline
 \hline
\end{tabular}
\label{tab1NNExt}
\end{table}

Therefore, despite the limitation to low levels, accurate estimates for the critical parameters (for the square lattice) can be obtained with this method. This is in agreement with the results by Monroe \cite{Monroe} for the ferromagnetic Ising model, where a critical temperature differing by 0.003\% from the Onsager value was obtained from extrapolations of data for square HLs for $L \leqslant 5$.

\section{Critical parameters for the 2NN model}
\label{secRes2NN}

Now, we investigate the 2NN model on square HLs of levels up to $L=7$ and $L=8$, respectively, in the approximations which overestimate (O) and underestimate (U) the number of NNNs for some sites. %The densities and free energies were calculated only in the O case, for $L\leqslant 6$.
Similarly to the 1NN model, here, in both approximations, the RRs assume a homogeneous solution, related to the fluid ($F$) phase, characterized by $R_{\sigma,A}=R_{\sigma,B}=R_{\sigma,C}=R_{\sigma,D}$, for $\sigma=1,...,N-1$, which is stable for small $z_2$. For large $z_2$, in both O and U cases, there are four equivalent fixed points associated with the columnar nature of the ordered phase of the 2NN model. Although these fixed points are not so simple as in the fluid phase or in the solid phase of the 1NN model, by inspection of the values of $R_{\sigma,S}$ and of the densities, it is easy to verify, for example, that there are two possible ways for the sublattice $A$ be more occupied: one in which the sublattice $B$ is also more occupied; and other with the sublattice $D$, instead of $B$, being more populated. Each of these fixed points are associated with columns being formed in one of the two directions of the square building blocks.

\begin{figure}[t]
 \includegraphics[width=8.5cm]{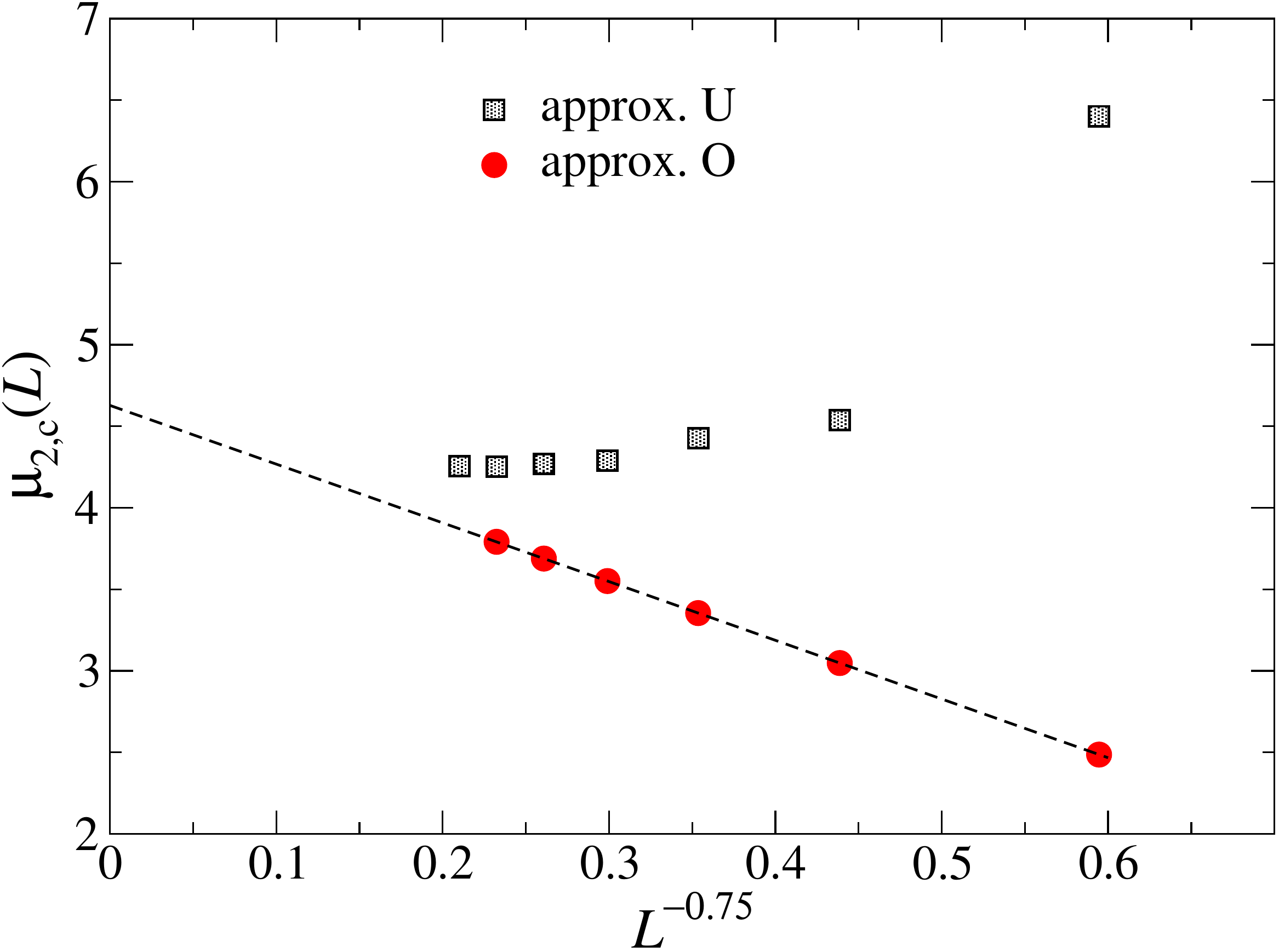}
 \caption{Comparison of $\mu_{2,c}(L)$ versus $L^{-0.75}$ for approximations U (black squares) and O (red circles). The dashed line is a linear fit.}
 \label{fig4}
\end{figure}

In all levels, the stability analysis of the fixed points, as well the behavior of particle densities, order parameters and free energies demonstrate that there exists a critical point, $\mu_{2,c}(L)$, separating the fluid and the columnar phases. This confirms that the fluid-columnar transition is continuous in the square lattice case, in agreement with most of the works on this hard-square model, as discussed in the Introduction. In Fig. \ref{fig4}, we compare the values of $\mu_{2,c}(L)$ for approximations O and U. In the former case, one observes a monotonic convergence as $L$ increases, whereas the results for approximation U display a nonmonotonic convergence, as well as a parity dependence on $L$. In fact, although this is not so clear in Fig. 4, one finds that $\mu^U_{2,c}(L=8)>\mu^U_{2,c}(L=7)$, indicating that for larger (and unfeasible) $L$'s, this chemical potential will pass to converge from below, similarly to case O. This is indeed expected, since for $L \rightarrow \infty$ both approximations shall give the same result. For small $L$'s, notwithstanding, the effect of underestimating the neighborhood of eight sites is very strong, hindering the particles' ordering, which explains why $\mu_{2,c}^U > \mu_{2,c}^O$, for a given $L$. For this reason, hereafter we will discuss only the results for approximation O, whose critical parameters $\mu_{2,c}$, $\rho_{2,c}$ and $\phi_{2,c}$ are shown in Tab. \ref{tab3}.

\begin{table}[b] \centering
\caption{Critical chemical potentials $\mu_{2,c}$, particle densities $\rho_{2,c}$ and free energies $\phi_{2,c}$ for the 2NN model on square HLs of different levels $L$, in approximation O.}
\begin{tabularx}{\columnwidth}{p{0.6cm} >{\centering\arraybackslash}X >{\centering\arraybackslash}X >{\centering\arraybackslash}X >{\centering\arraybackslash}X}%{l c c c c}
 \hline
 \hline
  $L$     & $\mu_{2,c}$   &    $\rho_{2,c}$     &   $\phi_{2,c}$   \\
  \hline
  $2$      & $2.486741$               &    $0.200419$   &     $0.519643$      \\
  $3$      & $3.048128$               &    $0.207922$   &     $0.457326$      \\
  $4$      & $3.354577$               &    $0.213566$   &     $0.393478$      \\
  $5$      & $3.550567$               &    $0.216899$   &     $0.340409$      \\
  $6$      & $3.688488$               &    $0.218782$   &     $0.300490$      \\
  $7$      & $3.791358$               &    ---           &     $0.267941$      \\
 \hline
 \hline
\end{tabularx}
\label{tab3}
\end{table}

It is noteworthy in Fig. \ref{fig4} that $\mu_{2,c}(L)$, for the case O, is well-linearized when plotted against $L^{-0.75}$, with a simple linear fit returning the extrapolated value $\mu_{2,c}(\infty) \approx 4.628$. This strongly indicates that $\mu_{2,c}(L)$ follows the finite-size scaling of Eq. \ref{zc} with $\Delta_1 = 3/4$. In fact, by assuming that $a_i=0$ for $i\geqslant 2$ in this equation and performing 3-pt extrapolations of the values of $\mu_{2,c}$ in Tab. \ref{tab3}, we obtain exponents quite close to $\Delta_1 = 3/4$, as shown in Tab. \ref{tab2NNExt}. The extrapolated values of the critical chemical potential are also displayed in Tab. \ref{tab2NNExt} and, in contrast to the 1NN case, they do not have a clear tendency to increase or decrease. This suggests that further finite-size corrections are very small in this quantity, as already hinted by the good linear behavior in Fig. \ref{fig4}. This is indeed confirmed by 4-pt extrapolations [considering $\Delta_1=3/4$, with $\mu_{2,c}(\infty)$, $a_1$, $a_2$ and $\Delta_2$ as unknowns in Eq. \ref{zc}], which yield $\Delta_2 \gtrsim 6$ for the largest $L$'s. Such extrapolations provides values in the range $4.626 \lessapprox \mu_{2,c}(\infty) \lessapprox 4.631$, once again, without any clear tendency to increase or decrease with $L$. Hence, we may regard $\mu_{2,c}^*=4.629(3)$ as our best estimate for the critical point of the 2NN model on the square lattice. Given the difficulties inherent to this model, it is remarkable that this value differs by $\lesssim 1$\% from several results from MC simulations (giving $\mu_{2,c} \approx 4.58$ \cite{Heitor,Zhitomirsky,Feng,Ramola2}), as well as from a recent interfacial tension calculation ($\mu_{2,c} \approx 4.66$ \cite{Rajesh2NN}).

\begin{table}[t] \centering
\caption{Results from 3-pt extrapolations of the critical parameters $\mu_{2,c}$, $\rho_{2,c}$ and $\phi_{2,c}$ in Tab. \ref{tab3}, considering Eq. \ref{zc}, for sets of levels $(L-1,L,L+1)$. The obtained exponents $\Delta_1$ from the extrapolations of $\mu_{2,c}$ are also shown.}
\begin{tabular}{l c c c c}
 \hline
 \hline
  Set of $L$'s      & \phantom{...}$\mu_{2,c}(\infty)$   &    $\Delta_1$      &   $\rho_{2,c}(\infty)$ &   $\phi_{2,c}(\infty)$  \\
  \hline
  $(2,3,4)$      & $4.62538$           &    $0.750954$    &  --         & $0.63410$  \\ 
  $(3,4,5)$      & $4.62265$           &    $0.752407$    &  $0.22935$  & $1.24463$  \\
  $(4,5,6)$      & $4.65061$           &    $0.734765$    &  $0.22360$  & $-0.2172$  \\
  $(5,6,7)$      & $4.63792$           &    $0.743947$    &  --         & $-0.6959$  \\
 \hline
 \hline
\end{tabular}
\label{tab2NNExt}
\end{table}

For calculating the critical density, one has to deal with generalized (and considerably enlarged) RRs, as explained in the Appendix, so that we were able to estimate this quantity only for $L \leqslant 6$. The outcomes from 3-pt extrapolations, assuming again the finite-size scaling of Eq. \ref{zc}, are displayed in Tab. \ref{tab2NNExt}. Similarly to the 1NN case, this extrapolation fails for the smallest set of sizes $(2,3,4)$, confirming that the densities for low-level HLs are indeed far from the asymptotic behavior. Anyhow, using such densities to perform 5-pt extrapolations (considering simple power-law corrections, as well as logarithmic ones) one always gets $\rho_{2,c} \approx 0.223$, in agreement with the extrapolated value in Tab. \ref{tab2NNExt} for the three largest $L$'s. This value is slightly smaller, but close to those reported in previous works $\rho_{2,c}\approx 0.233$ \cite{Heitor,Zhitomirsky}, with a difference of $\approx 4$\%. Although the value of $\rho_{2,c}(\infty)$ for the set $(4,5,6)$ is slightly smaller than the one for $(3,4,5)$ [see Tab. \ref{tab2NNExt}], given the fluctuations found in $\mu_{2,c}(\infty)$, this can not be seen as an indication that $\rho_{2,c}$ will converge to a value smaller than $0.233$.

Interestingly, the critical free energies found here for the 2NN model are positive (see Tab. \ref{tab3}), meaning that the critical pressures are negative in these systems. However, $\phi_{2,c}$ decreases fast with $L$ and our results strongly suggests that it converges to a negative value. This is indeed confirmed in Tab. \ref{tab2NNExt}, which shows results from 3-pt extrapolations (following Eq. \ref{zc}). The strong variation in $\phi_{2,c}(\infty)$ does not allow us to propose a value, not even approximated, for $\phi_{2,c}$ in the square lattice case. Moreover, unfortunately, 5-pt extrapolations fail in returning physical values in this case. This poor convergence certainly explains why results for $\phi_{2,c}$ are absent in the literature, to the best of our knowledge.

\section{Coherent anomaly method}
\label{secCAM}

\begin{figure}[t]
 \includegraphics[width=8.5cm]{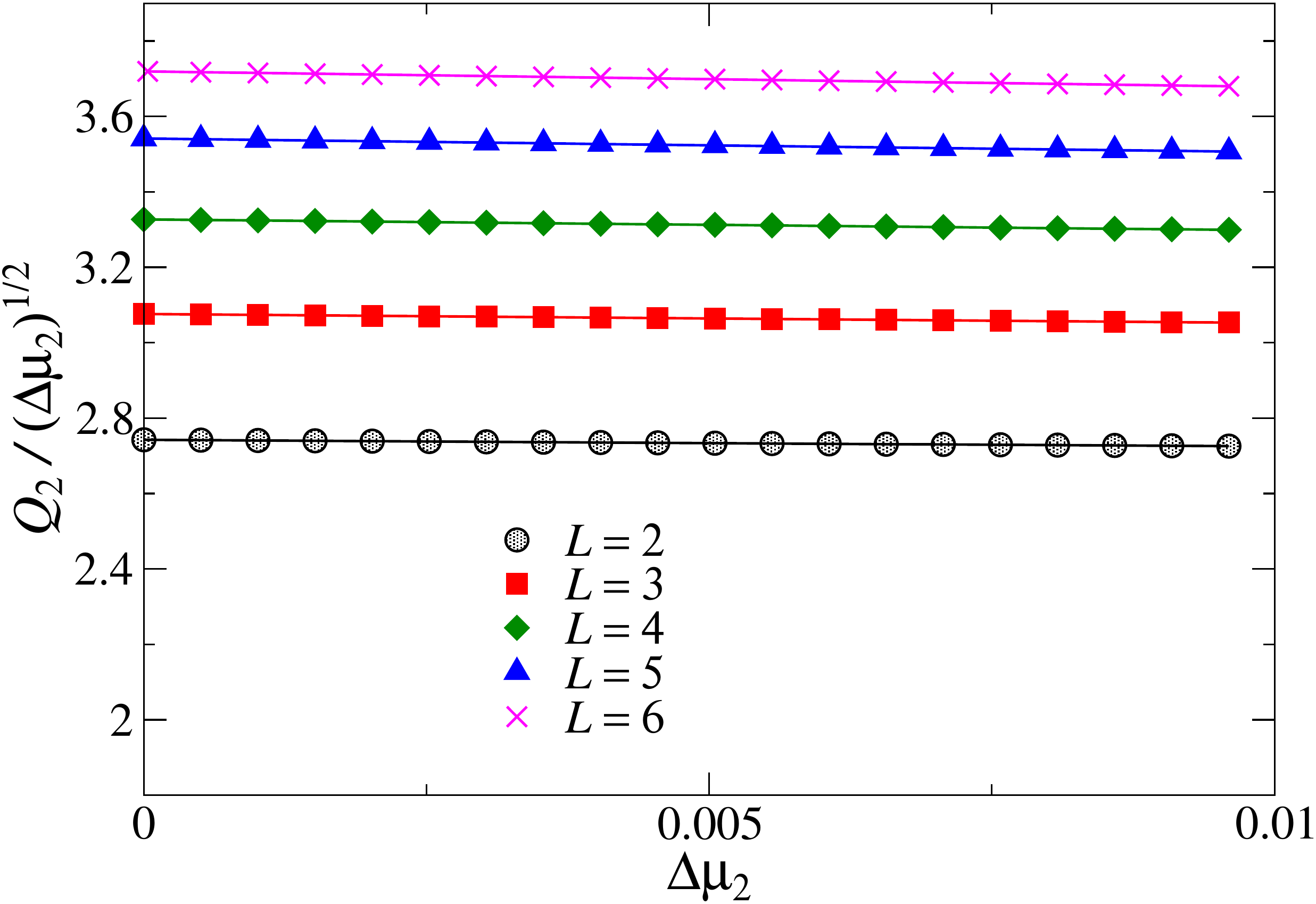}
 \caption{Rescaled order parameter $Q_2 / {(\Delta\mu_2)}^{\frac{1}{2}}$ versus $\Delta\mu_2$ for the 2NN model and several $L$'s, as indicated in the legend. The lines are linear fits used to extrapolate these data to $\Delta\mu_2 \rightarrow 0$.}
 \label{fig5}
\end{figure}

Next, we investigate the universality classes of both $k$NN models, on the square lattice, through the coherent anomaly method (CAM). We remark that, independently of its level $L$, the dimension of the square HLs is infinity. Hence, in all levels, the critical exponents assume their classical values in the continuous fluid-solid or fluid-columnar transitions discussed above. Of particular interest here will be the order parameters $Q_1$ and $Q_2$, defined in Eqs. \ref{eqQ1} and \ref{eqQ2}, respectively. Close to the critical point, they behave as
\begin{equation}
 Q_k(L) =\bar Q_k(L) \Delta \mu_k(L)^{\beta_{cl}},
 \label{eqQi}
\end{equation}
where $\bar Q_k(L)$ are non-universal amplitudes (the coherent anomalies), $\beta_{cl}=1/2$ is the classical critical exponent and $\Delta \mu_{k}(L) \equiv [\mu_{k}-\mu_{k,c}(L)]/\mu_{k,c}(L)$. Thereby, the amplitudes $\bar Q_k(L)$ can be estimated by extrapolating $Q_k / (\Delta \mu_k)^{\frac{1}{2}}$ for $\Delta \mu_k \rightarrow 0$, as done in Fig. \ref{fig5} for the 2NN model. A very similar behavior is found also in the 1NN case. These amplitudes are presented in Tab. \ref{tab4}.

According to the CAM theory \cite{SuzukiCAM}, the non-universal amplitudes of a sequence of systematically improved mean-field approximations, as is the case of our HL solutions \cite{Monroe}, encode information on the true critical exponents. For the order parameters, one has \cite{SuzukiCAM}
\begin{equation}
 \bar Q_k(L) = a \Delta\mu_k^{*}(L)^{\beta - 1/2},
 \label{eqCAM}
\end{equation}
where $a$ is a constant and
\begin{equation}
 \Delta \mu_k^*(L) \equiv \frac{\mu_{k,c}^*-\mu_{k,c}(L)}{\mu_{k,c}^*},
 \label{mu}
\end{equation}
with $\mu_{k,c}^*$ being the true critical point, for $L \rightarrow \infty$.

\begin{table}[!t] \centering
\caption{Scaling amplitudes $\bar Q_k$ of the order parameters [Eq. \ref{eqQi}], for several $L$'s, estimated from the extrapolations in Fig. \ref{fig5} for $k=2$, and analogous ones [not shown] for $k=1$.}
\begin{tabularx}{\columnwidth}{p{0.6cm} >{\centering\arraybackslash}X >{\centering\arraybackslash}X}
 \hline\hline
% \vspace{0.001cm}
  $L$      & $\bar{Q_1}$              & $\bar Q_2$    \\ 
  \hline
  $1$      & $1.190403$                      & ---                 \\
  $2$      & $2.976795$                      & $2.742649$            \\
  $3$      & $3.345415$                      & $3.076205$             \\
  $4$      & $3.652503$                      & $3.327170$             \\
  $5$      & $3.912666$                      & $3.541579$             \\
  $6$      & $4.142837$                      & $3.719521$             \\
 \hline\hline
\end{tabularx}
\label{tab4}
\end{table}

Therefore, at first, both $\beta$ and $\mu_{k,c}^*$ can be estimated from Eq. \ref{eqCAM}, by means of 3-pt extrapolations. This procedure has indeed been used with some success in several works (see, e.g., \cite{Monroe,XiaoHu,XiaoHu2,Hatori}). Particularly in the HL-based study of the ferromagnetic Ising model by Monroe \cite{Monroe}, critical temperatures differing by $\sim 0.1$\% from the Onsager value were found in extrapolations for different sets of levels, although with $\beta$ exponents $\approx 10$\% larger than the exact one. On the other hand, here, inaccurate results are found even for $\mu_{k,c}^*$ from such extrapolations. For instance, for the 1NN model one obtains $1.32 \lessapprox \mu_{1,c}^* \lessapprox 1.36$ and $0.07 \lessapprox \beta \lessapprox 0.13$, while in the 2NN case one gets $4.47 \lessapprox \mu_{2,c}^* \lessapprox 5.05$ and $0.0 \lessapprox \beta \lessapprox 0.18$.

In view of this, we adopt a different strategy, by letting $\mu_{k,c}^*$ fixed at the values estimated in the previous sections and calculating only the effective $\beta$ exponents through 2-pt extrapolations using Eq. \ref{eqCAM}. In the 1NN case, it does not matter whether one uses our value $\mu_{1,c}^* = 1.3340$ or the more accurate one from Ref. \cite{GuoBlote}, approximately the same exponents are found, being $\beta = 0.112$,  $0.120$ and $0.121$ for the pairs of levels $(3,4)$, $(4,5)$ and $(5,6)$, respectively. These estimates are increasing towards the expected value, $\beta=1/8$, with a deviation of 3\% observed in the result for the largest $L$'s.

In the 2NN case, one does not have a well-established value for $\mu_{2,c}^*$. For this reason, we employ the 2-pt extrapolations considering the value recurrently found in MC simulations $\mu_{2,c}^{*(1)} = 4.58$ \cite{Heitor,Zhitomirsky,Feng,Ramola2}, our estimate $\mu_{2,c}^{*(2)} = 4.629$ and the recent result from interfacial tension approach $\mu_{2,c}^{*(3)} = 4.66$ \cite{Rajesh2NN}. The obtained exponents are shown in Tab. \ref{tab6}. In all cases, no tendency to increase or decrease with $L$ is seen in the values of $\beta$, so, one may average them to obtain $\beta^{(1)}=0.146(14)$, $\beta^{(2)}=0.132(10)$ and $\beta^{(3)}=0.123(8)$. Thereby, by increasing $\mu_{2,c}^*$ by $\approx 1$\%, the exponent decreases by $\sim 10$\%. Its is noteworthy that $\beta^{(1)}$, estimated with the critical point from MC simulations, is considerably larger than the Ising exponent ($\beta_{Ising}=0.125$) and much larger than the Ashkin-Teller one ($\beta_{AT} \approx 0.115$) found in \cite{Ramola2}. The exponent $\beta^{(2)}$, obtained with our value for $\mu_{2,c}^*$, agrees with $\beta_{Ising}$ within the error bars and is $\approx 12$\% larger than $\beta_{AT}$. Finally, the exponent $\beta^{(3)}$ is quite close to $\beta_{Ising}$ and only 6\% larger than $\beta_{AT}$. Unfortunately, with this diversity of values, we are not in position to draw any conclusion about the universality class of the 2NN model, specially regarding a dispute between Ising and Ashkin-Teller criticality.

\begin{table}[!t] \centering
\caption{Effective critical exponents $\beta$ for the 2NN model, from 2-pt extrapolations considering Eq. \ref{eqCAM} with $\mu_{2,c}^{*(1)} = 4.58$, $\mu_{2,c}^{*(2)} = 4.629$ and $\mu_{2,c}^{*(3)} = 4.66$.}
\begin{tabularx}{\columnwidth}{p{1.3cm} >{\centering\arraybackslash}X >{\centering\arraybackslash}X >{\centering\arraybackslash}X}
 \hline\hline
Set of $L$'s & $\beta^{(1)}$ & $\beta^{(2)}$ & $\beta^{(3)}$ \\
\hline
(2,3)  &  0.133  &  0.122  &  0.116   \\
(3,4)  &  0.149  &  0.136  &  0.128   \\
(4,5)  &  0.142  &  0.126  &  0.116   \\
(5,6)  &  0.160  &  0.142  &  0.131   \\
 \hline\hline
\end{tabularx}
\label{tab6}
\end{table}

Usually, works on CAM bring an analysis of the true correlation length exponent $\nu$, through the distance between the pseudo-critical point and the true one (for $L \rightarrow \infty$), which in our variables can be written as
\begin{equation}
 \Delta \mu^*(L) \sim L^{-1/\nu}.
 \label{eqFSS}
\end{equation}
This relation follows from finite-size scaling (FSS) \cite{FisherFSS,*FisherFSS2}, where $L$ is the effective lateral size of the system. In our case, the HLs are always infinity, but even then improved values for the critical points are obtained as the effective lateral size of the building blocks ($\sim L$) increases, which somewhat justifies the use of $L$ in Eq. \ref{eqFSS}. In fact, due to their treelike structure, correlations are weakened along the HLs (when compared with the square lattice) and by increasing $L$ one might expect a proportional increase in the effective correlation length on these cacti.
%Moreover, its is easy to see that $N_s(L,M) = 2(2 L^2 + L)3^M - 2L^2$ is the total number of sites of a finite square HL with $M$ generations and level $L \geqslant 1$. Thereby, in the large $M$ limit, one has $N_s(L,M) \sim 4 L^2 (1 + \frac{1}{2L})3^M$, showing that the total number of sites in the HL is proportional to $L^2$, as expected and indicating again that $L$ can be a good measure for the effective linear size of these cacti.

To verify whether the FSS of Eq. \ref{eqFSS} is indeed valid in our approach, we start applying this analysis to the critical temperatures of the ferromagnetic Ising model on square HLs, reported in Tab. I of Ref. \cite{Monroe}. In this case, one expects that $\Delta T^*(L)\equiv T(L)-T^* \sim L^{-1/\nu}$, with $T^* = 2/\ln(1+\sqrt{2})$. Therefore, effective $\nu$ exponents can be estimated from 2-pt extrapolations, whose values are depicted in Tab. \ref{tab7}. Their large variation indicates that further corrections, beyond $1/L$, are very important in $(\Delta T^*)^\nu$. By assuming that such corrections have the form $a_1/x + a_2/x^2 + \cdots$, one obtains $\nu \approx 1.09$. Approximately the same result is found from 3-pt extrapolations of the $\nu$'s in Tab. \ref{tab7} for the largest $L$'s. This estimate, which is $9$\% larger than the Ising value ($\nu_{Ising}=1$), strongly indicates that FSS holds in our system with the true $\nu$ exponent, once the mean-field one is $\nu_{cl}=1/2$.

\begin{table}[!t] \centering
\caption{Effective critical exponents $\nu$ from 2-pt extrapolations considering Eq. \ref{eqFSS} (and its analogous for the temperature in the Ising case), for the Ising \cite{Monroe}, 1NN and 2NN models on square HLs of different levels.}
\begin{tabularx}{\columnwidth}{p{1.3cm} >{\centering\arraybackslash}X >{\centering\arraybackslash}X >{\centering\arraybackslash}X}
 \hline\hline
Set of $L$'s & $\nu$ (Ising) & $\nu$ (1NN) & $\nu$ (2NN) \\
\hline
(1,2)   &   1.329   &   1.451   &   ---    \\
(2,3)   &   1.221   &   1.335   &   1.334  \\
(3,4)   &   1.181   &   1.270   &   1.335  \\
(4,5)   &   1.160   &   1.232   &   1.335  \\
(5,6)   &    ---    &   1.208   &   1.332  \\
(6,7)   &    ---    &    ---    &   1.331  \\
 \hline\hline
\end{tabularx}
\label{tab7}
\end{table}

A similar conclusion is obtained for the 1NN model. In fact, the careful analysis from Sec. \ref{secRes1NN} provided strong evidence of a correction exponent $\Delta_1 = 1$ in Eq. \ref{zc} for the critical chemical potential. So, by considering that $\Delta_1 = 1/\nu$ in this case, we are lead to conclude that $\nu \approx 1$ for the 1NN model. Additional confirmation of this is obtained here, through the effective $\nu$ exponents calculated from 2-pt extrapolations (using Eq. \ref{eqFSS} with $\mu_{1,c}^* = 1.3340151002$ \cite{GuoBlote}), which are displayed in Tab. \ref{tab7}. In fact, a 3-pt extrapolation of $1/\nu$ to $L \rightarrow \infty$, for the largest $L$'s, yields $\nu \approx 1.06$.

The exponents obtained for the 2NN model, with $\mu_{2,c}^* = 4.629$, are also shown in Tab. \ref{tab7}. In this case, they are always quite close to $\nu \approx 1.33$, which is consistent with the correction exponent $\Delta_1 =1/\nu \approx 0.75$ found in Sec. \ref{secRes2NN}. We remark that for  $\mu_{2,c}^* = 4.58$ ($\mu_{2,c}^* = 4.66$) one finds exponents with a tendency to decrease (increase), which extrapolate to $\nu \approx 1.22$ ($\nu \approx 1.37$), being $\approx 8$\% smaller ($\approx 3$\% larger) than $1.33$. Hence, for all values of $\mu_{2,c}$ the exponents are considerably larger than $\nu_{Ising}$ and the Ashkin-Teller exponent found in MC simulations: $\nu_{AT} \approx 0.92$  \cite{Ramola2}. We recall also that $\beta/\nu=1/8$ in both Ising and Ashkin-Teller criticality, while our exponents give $\beta/\nu \approx 0.12$, $\beta/\nu \approx 0.10$ and $\beta/\nu \approx 0.09$, respectively for $\mu_{2,c}^* = 4.58$, $\mu_{2,c}^* = 4.629$ and $\mu_{2,c}^* = 4.66$. Therefore, if one assumes that the correct critical exponents for the 2NN model are the Ashkin-Teller ones from \cite{Ramola2}, the critical chemical potential from MC simulations returns the most accurate exponent ratio $\beta/\nu$, while it gives the worse estimate for $\beta$. Conversely, with $\mu_{2,c}^* = 4.66$ \cite{Rajesh2NN} one obtains the best result for $\beta$, but the largest deviation in $\nu$ and $\beta/\nu$.

\section{Final discussions and conclusion}
\label{secConc}

We have presented semi-analytical solutions of athermal $k$NN models, for $k=1$ and $k=2$, defined on Husimi lattices built with diagonal square lattices, with $2L(L+1)$ sites. For all $L$'s considered, the $k$NN models exhibit thermodynamic behaviors analogous to those observed in the square lattice, with a continuous fluid-solid transition in the 1NN case and a continuous fluid-columnar transition in the $2\times 2$ hard-square (2NN) model. By increasing $L$, a systematic sequence of even better values for the critical parameters was obtained in the 1NN case, as well as for the 2NN model in an approximation which overestimate the neighborhood of some few sites. (A second approximation for the 2NN model, underestimating this neighborhood, proved to be non-systematic for the $L$'s analyzed here.) With this method, at one hand, it is quite hard to study large $L$'s, once the computational resources (both HD space and RAM memory) needed to handle the very large number of terms in the recursion relations increase exponentially with $L$. On the other hand, extrapolations of the critical parameters obtained for low levels (to $L \rightarrow \infty$) yield results in quite good agreement with the best available estimates for them in the square lattice. For instance, for the 1NN model this gives a critical chemical potential differing by 0.001\% from the high accurate result from transfer-matrix reported in \cite{GuoBlote}. For the 2NN model, our value is $\approx 1$\% larger than those typically found in MC studies \cite{Heitor,Zhitomirsky,Feng,Ramola2}. Given the previous results for this last model, with far diverse values for $\mu_{2,c}$ obtained with different approaches, it is quite impressive that our method (with $L \leqslant 7$) furnishes a result so close to the one from large scale simulations. This certainly happens because, at each level, we calculate (with high precision) the \textit{true} critical parameters for \textit{infinite} HLs, while other approaches provide \textit{pseudo}-critical estimates for \textit{finite} (and usually small) regular lattices. This indicates that solutions on generalized HLs are indeed a very effective way to access the quantitatively correct phase behavior of lattice models in general, although it may be difficult to investigate systems with long-range interactions with this method, because of the issue with the definition of high-order neighbors in these trees.

Using the coherent anomaly method (CAM) for the order parameters and for the shifts $\mu_{k,c}^* - \mu_{k,c}(L)$, we have estimated the critical exponents $\beta$ and $\nu$ (for the square lattice). Our results for the 1NN model [$\beta \approx 0.121$ and $\nu \approx 1.06$] are close to the expected Ising exponents. In the 2NN case, however, small changes in the value of the asymptotic critical potential $\mu_{2,c}^*$ used to calculate $\beta$ and $\nu$ lead to considerable variations in these exponents. It is very interesting that with $\mu_{2,c}^* = 4.629$ (as estimated here) one obtains $\beta = 0.132(10)$ and $\nu \approx 1.33$, which agree quite well with the exponents for ordinary percolation in two-dimensions ($\beta = 5/36$ and $\nu=4/3$). It turns out however that the four-fold symmetry breaking in the fluid-columnar transition of the 2NN model has no clear relation with percolation and, thus, this agreement seems to be a simple coincidence. In fact, the Ashkin-Teller criticality with $\nu \approx 0.92$ and $\beta/\nu = 1/8$, suggested in previous numerical works \cite{Feng,Ramola2}, is a much more plausible scenario. Our estimates for $\beta$ are indeed not so incompatible with this. The high positive deviation in $\nu$ (of $\approx 45$\%) may be caused by the overestimation in the number of second neighbors of some sites of the HLs. However, similar deviations in $\nu$ have been observed in a previous CAM study \cite{Patrykiejew} for the Ising model and soft lattice gas systems, using MC cluster approximations, where the problem with the definition of second neighbors is absent. This suggests that the large $\nu$ found here is more likely a failure of the CAM analysis for low levels than an indication of another universality class for the 2NN model.

\acknowledgments

We acknowledge financial support from CNPq, CAPES and FAPEMIG (Brazilian agencies) and the use of the Computing Cluster of the Universidade Federal de Vi\c cosa. We thank I. S. S. Carrasco for helpful discussions and J. F. Stilck for a critical reading of the manuscript.

\appendix

\section{Solution of the $k$NN models on the square Husimi lattices}
\label{secApend}

\begin{figure}[b]
 \includegraphics[width=8.5cm]{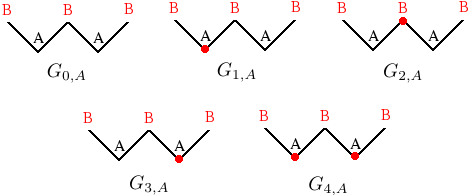}
 \caption{Possible states for the zigzag root lines of HLs of level $L=2$, for the 1NN model. The red dots indicate the 1NN particles. Another set of identical configurations exists for the root line where the sublattices $A$ and $B$ are exchanged, which defines the ppf's $G_{\sigma,B}$, for $\sigma =0,\ldots,4$.}
 \label{fig6}
\end{figure}

To solve the $k$NN models on a $L$-level square Husimi lattice, we define partial partition functions (ppf's) associated with the possible states of a root zigzag line [see Fig. \ref{fig3}] of a rooted building block (RBB). For the 1NN model, these root lines can be defined with $2L-1$ sites, whereas to include exclusion among external next-nearest neighbors in the 2NN case they must have $2L+1$ sites [see the definitions in Fig. \ref{fig3}]. Let us focus on the former case, since the extension to the latter one is immediate. In the 1NN model with $L=2$, for example, there are $N=5$ configurations of particles for the root line, as shown in Fig. \ref{fig6}, corresponding to all sites empty ($\sigma=0$), one site occupied (there are three possibilities here, $\sigma=1,2,3$) and two sites occupied ($\sigma=4$). Since these sites can be in two different sublattice configurations (the sequences $ABA$ or $BAB$), there is a total of $10$ possible states for the root line of the 1NN model with $L=2$. For the sake of simplicity, we will use only the sublattice of the leftmost site of the root line to identify the sublattice configuration; namely, we will use $A$ to denote $ABA$ and $B$ to $BAB$. Thereby, in this case one has $10$ ppf's: $G_{\sigma,S}$, with $\sigma=0,\ldots,4$ and $S=A,B$. For comparison, for the 2NN model with $L=2$ one has $N=9$ and four sublattice configurations, totaling $36$ states and ppf's: $G_{\sigma,S}$, with $\sigma=0,\ldots,8$ and $S=A,\ldots,D$. In our solutions, we will always define the configuration $\sigma=0$ as the one with the root line empty. In Tab. \ref{tab1} the total number of ppf's is presented for different levels, for both models.

\begin{table}[!t] \centering
\caption{Total number of ppf's ($2N$ and $4N$, respectively) and total number of terms $K_T$ in the RRs for the ppf's of the 1NN (top) and 2NN (bottom) models on $L$-level square HLs.}
\label{tab1}
\begin{tabularx}{\columnwidth}{p{1.3cm} >{\centering\arraybackslash}X >{\centering\arraybackslash}X}
\hline
\hline
  $L$       & $2 N$ (1NN)    &  $K_T$ (1NN)                      \\
\hline
  $2$       & $10$         &  $1704$                        \\
  $3$       & $26$         &  $1146118$                     \\
  $4$       & $68$         &  $3985772648$                  \\
  $5$       & $178$        &  $70897617428720$              \\
  $6$       & $466$        &  $6438412592897497526$         \\
%  $7$     &  & $1220$      & &  --                            \\
\hline
  $L$       & $4 N$ (2NN)    & $K_T$ (2NN)                      \\ 
\hline
  $2$       & $36$         & $8312$                         \\
  $3$       & $76$         & $950636$                       \\
  $4$       & $164$        & $357340560$                    \\
  $5$       & $352$        & $435265986532$                 \\
  $6$       & $756$        & $258990287426480$              \\
  $7$       & $1624$       & $3729631034070503744$          \\
\hline
\hline
\end{tabularx}
\end{table}

\begin{figure}[t]
 \includegraphics[width=8.5cm]{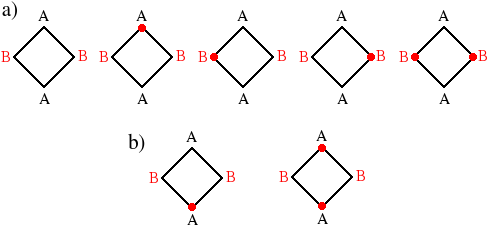}
 \caption{Possible configurations for the rooted square of HLs of level $L=1$, for the 1NN model, when the root site is in sublattice $A$ and in the states a) $\sigma=0$ and b) $\sigma = 1$. The red dots indicate the 1NN particles.}
 \label{fig7}
\end{figure}

A recursion relation (RR) for the ppf $G_{\sigma,S}$ can be obtained by keeping a RBB with the root line in the state ($\sigma,S$) and considering the operation of attaching three subtrees (three branches) to it, one at each of its sides, with exception of the root line. If each of these subtrees has $M$ generations, this process yields a new subtree with $M+1$ generations. By summing over all possible ways of attaching the three subtrees to the RBB --- i.e., by considering all the possible configurations for their root lines, respecting the particle exclusions and sublattice order ---, one obtains the ppf $G'_{\sigma,S}$ in generation $M+1$ as a polynomial function of the ppf's $G_{i,J}$ in generation $M$. For example, for the 1NN model on the ordinary square HL (i.e., the $L=1$ case), the possible configurations for the RBB, when the root site is in sublattice $A$, are depicted in Fig. \ref{fig7}. They yield the RRs:
\begin{subequations}
\label{eqGisL1}
\begin{eqnarray}
 G'_{0,A} = G_{0,A}G_{0,B}^2 &+& z_{1}^{\frac{1}{2}} G_{1,A}G_{0,B}^2 \\ \nonumber
 &+& 2 z_{1}^{\frac{1}{2}} G_{0,A}G_{0,B}G_{1,B} + z_{1} G_{0,A} G_{1,B}^2
\end{eqnarray}
and
\begin{equation}
 G'_{1,A} = z_1^{\frac{1}{2}} \left[ G_{0,A}G_{0,B}^2 + z_{1}^{\frac{1}{2}} G_{1,A}G_{0,B}^2 \right].
\end{equation}
\end{subequations}
The RRs for $G_{0,B}$ and $G_{1,B}$ are given by the same expressions with $A$ and $B$ exchanged.

Similarly to Fig. \ref{fig7}, if the root line is in sublattice configuration $A$ for the 1NN model when $L>1$, then, one has to connect a subtree in configuration $A$ at the top of the RBB and subtrees in configuration $B$ at its lateral sides [see Fig. \ref{fig1}(c)]. This means that the RRs for the ppf's $G'_{\sigma,A}$ will be always given by a sum of terms of the form $G_{\alpha,B}G_{\beta,A}G_{\gamma,B}$, with $\alpha,\beta,\gamma=0,\ldots,N-1$. We notice that there exits several combinations of $G_{\alpha,B}G_{\beta,A}G_{\gamma,B}$ which are forbidden, due to the particle exclusions, so that the number of allowed terms in the RRs is smaller than $N^3$, as it is clear in Eqs. \ref{eqGisL1} (for which $N=2$). Anyhow, we can sum over all $N^3$ configurations --- i.e., over all possible sets $\{\alpha,\beta,\gamma\}$ --- and introduce a variable $\delta_{\sigma;\alpha,\beta,\gamma}$, such that $\delta_{\sigma;\alpha,\beta,\gamma}=0$ for the forbidden configurations and $\delta_{\sigma;\alpha,\beta,\gamma}=1$ otherwise. In this way, the RRs for $G_{\sigma,A}$, for the 1NN model, and general $L$ can be written as 
\begin{equation}
\label{eqGis}
 G'_{\sigma,A}=\sum_{\{\alpha,\beta,\gamma\}} \delta_{\sigma;\alpha,\beta,\gamma} z_{1}^{\frac{1}{2}n_{\sigma;\alpha,\beta,\gamma}} f_{\sigma;\alpha,\beta,\gamma}(z_{1}) G_{\alpha,B} G_{\beta,A} G_{\gamma,B},
\end{equation}
where $n_{\sigma;\alpha,\beta,\gamma}$ is the number of particles at the $4L$ more external sites of the building blocks, which are shared by two consecutive generations of the tree. Because of this, $n_{\sigma;\alpha,\beta,\gamma}$ appears multiplied by half. The contribution of the particles belonging exclusively to the RBB is accounted in the polynomial
\begin{equation}    
f_{\sigma;\alpha,\beta,\gamma}(z_{1}) = \sum^{K_{\sigma;\alpha,\beta,\gamma}}_{i=k_\sigma} m_{i;\sigma;\alpha,\beta,\gamma} z_{1}^{i},
 \label{eqfGi}
\end{equation}
where $k_\sigma$ [$K_{\sigma;\alpha,\beta,\gamma}$] is the minimal [maximal] number of particles that can be placed in the bulk sites of the RBB. Note that the state $\sigma$ of the root line fixes the configuration of some bulk sites of the RBB, such that not necessarily $k_\sigma = 0$ in $f_{\sigma;\alpha,\beta,\gamma}$. In this function, $m_{i;\sigma;\alpha,\beta,\gamma}$ gives the number of ways of placing $i$ particles in the bulk sites, with $k_\sigma$ of them fixed. In the $L=1$ case, where there is no bulk sites in the RBB, one has $f_{\sigma;\alpha,\beta,\gamma}=1$. The RRs for the ppf's $G'_{\sigma,B}$ are given by Eq. \ref{eqGis} with $A$ and $B$ exchanged.

For the 2NN model, beyond replacing $z_1$ by $z_2$ in Eqs. \ref{eqGis} and \ref{eqfGi}, the former equation has to be generalized to four sublattices. For example, for even $L$, when the root line is in sublattice configuration $A$, as it is the case in Fig. \ref{fig1}(d), subtrees in configurations $B$ and $D$ will be attached at the lateral sides of the RBB, while at the top side the incoming subtree shall be in configuration $A$. Hence, we must have $G_{\alpha,B} G_{\beta,A} G_{\gamma,D}$ in place of $G_{\alpha,B} G_{\beta,A} G_{\gamma,B}$ in Eq. \ref{eqGis}. When $L$ is odd, we will have $G_{\alpha,B} G_{\beta,C} G_{\gamma,D}$ in Eq. \ref{eqGis}, since in this case the subtree attaching at the top of the RBB shall be in configuration $C$. Once the RRs for ppf's $G'_{\sigma,A}$ are determinated, the ones for the other sublattices can be obtained by cyclic permutations of their indexes: $A \rightarrow B$, $B \rightarrow C$, $C \rightarrow D$ and $D \rightarrow A$.

To determine the set of integers $\delta_{\sigma;\alpha,\beta,\gamma}$, $n_{\sigma;\alpha,\beta,\gamma}$, $m_{i;\sigma;\alpha,\beta,\gamma}$, $k_\sigma$ and $K_{\sigma;\alpha,\beta,\gamma}$ in Eqs. \ref{eqGis} and \ref{eqfGi}, we use an exact enumeration process. For small $L$'s this procedure can be easily done, but it becomes very computationally demanding as $L$ increases. To illustrate the complexity of this method, the sum of the number of allowed terms in all ppf's for a given $L$ [$K_T = \sum_{\{\sigma,\alpha,\beta,\gamma\}} \delta_{\sigma;\alpha,\beta,\gamma} (K_{\sigma;\alpha,\beta,\gamma}+1-k_\sigma)$] is shown in Tab. \ref{tab1} for both models. This number becomes $\sim 10^{18}$ already for $L=6$ ($L=7$) for the 1NN (2NN) model. In the 2NN case these values are for the approximation O, where the neighborhood of some sites is overestimated. To efficiently enumerate these large amount of configurations, we use a recursive method, where the information for $L-1$ is used to determine the quantities for the ppf's of the $L$th level. A caveat of the method is that, to save time, all informations of all RRs for the case $L-1$ have to be stored in the RAM memory, to allow rapid access. For instance, these informations for $L=8$ in the 2NN case would require at least $256$GB of RAM memory, while in HD they would occupy $\approx 5$ terabytes when the data are compressed!

After obtaining the ppf's for a given model and level, we start the study of its thermodynamic properties. In the thermodynamic limit, which corresponds to infinite HLs ($M \rightarrow \infty$), the RRs for the ppf's diverge. So, we work with ratios of them, which are defined here as $R_{\sigma,S}=\frac{G_{\sigma,S}}{G_{0,S}}$, with $\sigma=1,\ldots,N-1$ and $S=A,B$ ($S=A,\ldots,D$) in the 1NN (2NN) case. Recursion relations for these ratios can be easily obtained from Eqs. \ref{eqGis} and \ref{eqfGi}. For instance, for the 1NN model with $L=1$, they read
\begin{subequations}
\label{eqRRsL1}
\begin{eqnarray}
 R'_{1,A} = \frac{z_1^{\frac{1}{2}} + z_{1} R_{1,A}}{1 + z_{1}^{\frac{1}{2}} R_{1,A} + 2 z_{1}^{\frac{1}{2}} R_{1,B} + z_{1} R_{1,B}^2}
\end{eqnarray}
and
\begin{equation}
 R'_{1,B} = \frac{z_1^{\frac{1}{2}} + z_{1} R_{1,B}}{1 + z_{1}^{\frac{1}{2}} R_{1,B} + 2 z_{1}^{\frac{1}{2}} R_{1,A} + z_{1} R_{1,A}^2}.
\end{equation}
\end{subequations}

The real and positive fixed points of these RRs correspond to the phases of the model on the HL. As discussed in Sec. \ref{secRes1NN} (\ref{secRes2NN}), for the 1NN (2NN) model one has three (five) of such fixed points, being one associated with the fluid phase and the other ones with the equivalent configurations of the ordered solid (columnar) phase. The region where a given phase is stable is determined by the condition $\Lambda \leqslant 1$, being $\Lambda$ the maximum eigenvalue of the Jacobian matrix for the RRs of the ratios applied in the corresponding fixed point. The stability limit (i.e., the spinodal) of the phase is given by $\Lambda=1$. In all levels and for both models, the disordered and ordered phases are stable respectively for small and large $z_k$, and their spinodals coincide at critical activities $z_{k,c}(L)$. Therefore, in each level, the 1NN (2NN) model undergoes a continuous fluid-solid (fluid-columnar) transition.

Similarly to the ppf's, the partition function, $Y_k$, of a given $k$NN model on a $L$-level HL can be obtained by summing over all possible ways of attaching four subtrees with $M \rightarrow \infty$ generations to a central building block. In general, it can be written as
\begin{equation}
 Y_k = \sum^{N-1}_{i=0}\sum^{N-1}_{j=0}\delta'_{ij}z_{k}^{\frac{1}{2}n_j} G_{j,A}G'_{i,A},
 \label{eqY}
\end{equation}
where $\delta'_{ij}=1$ if the particle configuration of $G_{j,A}$ matches that of $G'_{i,A}$ at the $L$ shared sites of the root line and respect the particle exclusions; and $\delta'_{ij}=0$ otherwise. The number of particles in such shared sites is given by $n_j$. In the 1NN case, one may write $Y_1 = G_{0,A}^2 G_{0,B}^2 y_1$, while for the 2NN model one has $Y_2 = G_{0,A}^2 G_{0,B} G_{0,D} y_2$ or $Y_2 = G_{0,A} G_{0,B} G_{0,C} G_{0,D} y_2$ depending on whether $L$ is even or odd. In all cases, the functions $y_k$ depend only on the ratios and activities. 

Following the ansatz proposed by Gujrati \cite{Gujrati}, the (reduced) free energy \textit{per site} at the central building block of the HL reads
\begin{equation}
 \phi_k= -\frac{1}{2V_{eff}}\ln\left(\frac{\prod_i \Omega_{0,i}}{y_k^2}\right),
\end{equation}
where $V_{eff} = 2L^2$ and $\Omega_{0,i}$ is given by
\begin{equation}
 \Omega_{0,i}=\frac{G'_{0,i}}{\prod_j G_{0,j}},
\end{equation}
with $i,j=A,B$ for the 1NN model and $i,j=A,B,C,D$ in the 2NN case. 

To calculate the particle densities in each sublattice at the four sites of the central \textit{plaquette} of the central building block [see Fig. \ref{fig3}], we use a trick where generalized ppf's and RRs are defined by associating activities $z_{kS}$ to the particles in sublattice $S$ in these four sites, and $z_k$ to the rest. In this way, the polynomial in Eq. \ref{eqfGi} becomes $f_{\sigma;\alpha,\beta,\gamma}(z_1,z_{1A},z_{1B})$ in the 1NN model and $f_{\sigma;\alpha,\beta,\gamma}(z_2,z_{2A},z_{2B},z_{2C},z_{2D})$ in the 2NN case. Since the partition functions will be also functions of $z_{kS}$, the densities $\rho_{kS}$ can be determined as
\begin{equation}
 \rho_{kS}=\left.\frac{z_{kS}}{4Y_k}\frac{\partial Y_k}{\partial z_{kS}}\right|_{z_{kS}=z_k}.
 \label{eqRho}
\end{equation}

\end{document}